\documentclass[prl,twocolumn,amsmath,aps,superscriptaddress,letterpaper,tightenlines]{revtex4}
\usepackage{graphicx,xspace,units}
\usepackage{float}
\usepackage{amsmath}
\usepackage{amssymb}
\usepackage[normalem]{ulem}
\usepackage{wrapfig}
\usepackage[dvipsnames,usenames]{color}

\newcommand{\outline}[1]{}

\newcommand{\refeq}[1]{(\ref{eq:#1})}
\newcommand{\refeqn}[1]{Eq.~(\ref{eq:#1})}
\newcommand{\labeleqn}[1]{\label{eq:#1}}
\newcommand{\reffig}[1]{Fig.~\ref{fig:#1}}
\newcommand{\be}{\begin{equation}}
\newcommand{\ee}{\end{equation}}

\newcommand{\curl}{\boldsymbol{\nabla} \times}

\newcommand{\tr}{\text{tr}}

\newcommand{\T}{\mathbb{T}}

\newcommand{\X}{\mathbb{X}}

\newcommand{\tG}{\mathbb{G}}

\newcommand{\vecx}{\mathbf{x}}

 \newcommand{\veck}{\mathbf{k}}
\newcommand{\vecd}{\mathbf{d}}

\newcommand{\ep}{\epsilon}
\newcommand{\ka}{\kappa}
\newcommand{\I}{\mathbb{I}}
\newcommand{\V}{\mathbb{V}}
\newcommand{\bnabla}{\boldsymbol{\nabla}}

\begin{document}

\title{Constraints on stable equilibria with fluctuation-induced forces}
\author{Sahand Jamal Rahi}
\affiliation{Massachusetts Institute of Technology, Department of
  Physics, Cambridge, Massachusetts 02139, USA}
\author{Mehran Kardar}
\affiliation{Massachusetts Institute of Technology, Department of
  Physics, Cambridge, Massachusetts 02139, USA}
\author{Thorsten Emig}
\affiliation{Institut f\"ur Theoretische Physik, Universit\"at zu
  K\"oln, Z\"ulpicher Strasse 77, 50937 K\"oln, Germany}
\affiliation{Laboratoire de Physique Th\'eorique et Mod\`eles
  Statistiques, CNRS UMR 8626, B\^at.~100, Universit\'e Paris-Sud, 91405
  Orsay cedex, France}

\begin{abstract}
We examine whether fluctuation-induced forces can lead to stable levitation.  
First, we analyze a collection of classical objects at finite temperature that contain 
fixed {\em and} mobile charges, and show that any arrangement in space is unstable to small perturbations in position.  
This extends Earnshaw's theorem for electrostatics by including thermal fluctuations of internal charges.  Quantum fluctuations of the electromagnetic field are responsible for Casimir/van der Waals interactions.  Neglecting permeabilities, we find that any equilibrium position of items subject to such forces is also unstable if the permittivities of {\it all} objects are higher or lower than that of the enveloping medium; the former being the generic case for ordinary materials in vacuum.

\end{abstract}
\maketitle

Earnshaw's theorem~\cite{Earnshaw42} states that a collection of charges 
cannot be held in stable equilibrium solely by electrostatic forces. 
The charges can attract or repel, but cannot be stably levitated. 
While the stability of matter (due to quantum phenomena), and dramatic demonstrations
of levitating frogs~\cite{Geim98}, are vivid reminders of the caveats to this theorem,
it remains a powerful indicator of the constraints to stability in electrostatics.
There is much current interest in forces induced by fluctuating charges (e.g., mobile ions in solution),
or fluctuating electromagnetic (EM) fields (e.g., the Casimir force between metal plates).
The former (due to thermal fluctuations) may lead to unexpected phenomena such as 
attraction of like-charged macroions, and is thought to be relevant to interactions of biological molecules.
The latter (due mainly to quantum fluctuations) is important to the attraction (and stiction) of
components of microelectromechanical (MEM) devices. 
Here, we  extend Earnshaw's theorem to some fluctuation-induced forces,
thus placing strong constraints on the possibility of obtaining stable equilibria, and repulsion 
between neutral objects.

An extension of Earnshaw's theorem~\cite{Earnshaw42} to polarizable objects by Braunbek~\cite{Braunbek39-1,Braunbek39-2} establishes that dielectric and paramagnetic ($\ep>1$ and $\mu>1$) matter cannot be stably levitated by electrostatic forces, while diamagnetic ($\mu<1$) matter can. 
This is impressively demonstrated by superconductors and frogs that fly freely above magnets~\cite{Geim98}. 
If the enveloping medium is not vacuum, the criteria for stability are modified by substituting the static electric permittivity $\ep_M$ and magnetic permeability $\mu_M$ of the medium in place of the vacuum value of $1$ in the respective inequalities. 
In fact, if the medium itself has a dielectric constant higher than the objects ($\ep<\ep_M$), stable levitation  is possible, as demonstrated  for bubbles in liquids (see Ref.~\cite{Jones95}, and references therein).
For dynamic fields the restrictions of electrostatics do not apply; for example, lasers can lift and hold dielectric beads with index of refraction $n=\sqrt{\ep\mu}>1$~\cite{Ashkin70}.
In addition to the force which keeps the bead in the center of the laser beam, there is radiation pressure which pushes the bead along the direction of the Poynting vector. 
Ashkin and Gordon have proved that no arrangement of lasers can stably levitate an object just based on radiation pressure~\cite{Ashkin83}.

\begin{figure}[htbp]
\includegraphics[width=1.0\linewidth]{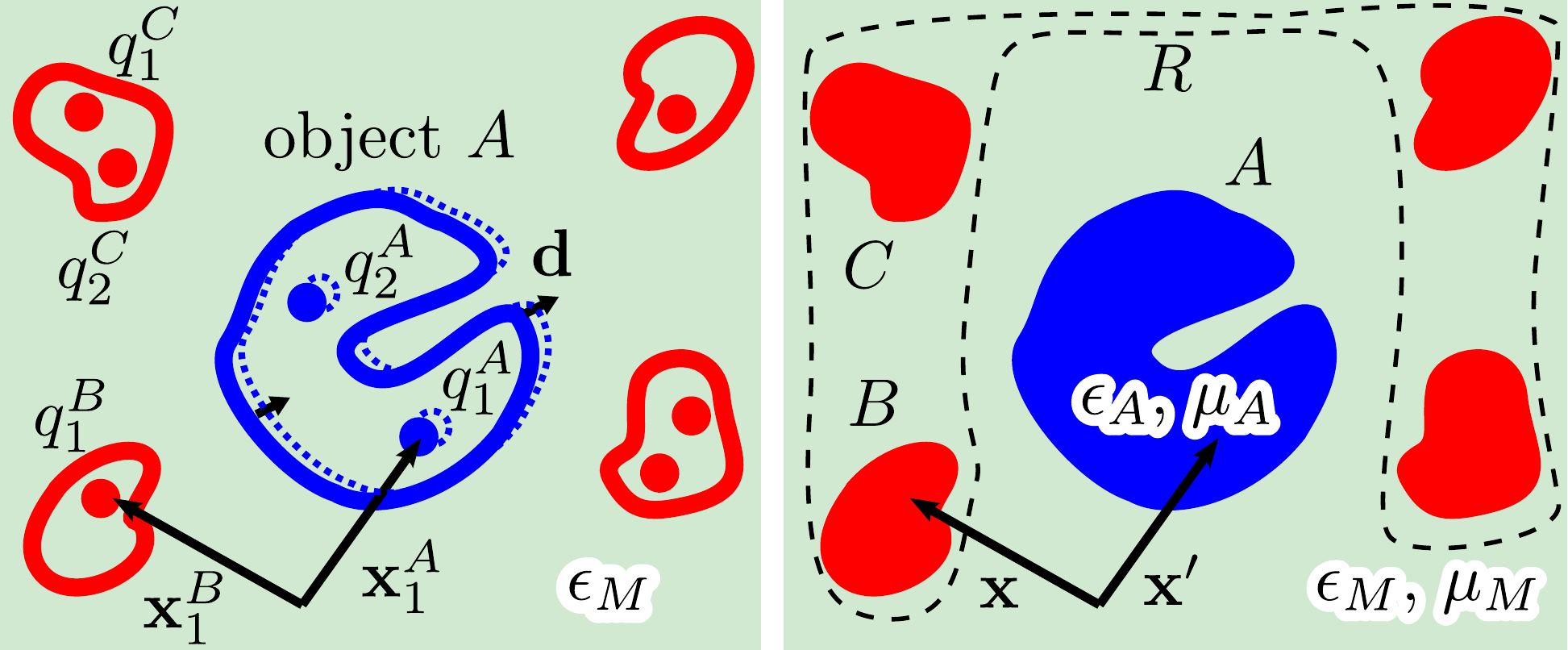}
\caption{(Color online) {\bf Left:} Each object contains mobile and stationary charges, which interact with charges in other objects according to Coulomb's law. 
They also interact with charges in the same object in an arbitrary manner and may be subject to an object-centered one-body potential. 
The medium has static permittivity $\epsilon_M$. The stability of the position of object $A$ is probed by displacing it infinitesimally by vector $\vecd$. 
{\bf Right:} The Casimir energy is considered for objects with electric permittivity $\epsilon_i(\omega,\vecx)$ and magnetic permeability $\mu_i(\omega,\vecx)$, embedded in a 
medium with uniform, isotropic, $\epsilon_M(\omega)$ and $\mu_M(\omega)$. 
To study the stability of object $A$, the rest of the objects are grouped in the combined entity $R$.}
\label{fig:all}
\end{figure}

We first obtain a simple extension of Earnshaw's theorem to
 objects containing fixed {\em and mobile} charges interacting  via Coulomb forces. 
This model, depicted in \reffig{all} (left), is 
a classical analogue of the electrodynamic Casimir effect, where quantum fluctuations produce spontaneous multipole moments and currents. 
Does Earnshaw's theorem, which states that arrangements of  \emph{fixed} charges is not stable to perturbations, also apply to  objects containing mobile charges?
The free energy is obtained, via the partition function, by integrating the positions $\{\vecx_i^J\}$ of the {\em mobile} charges $\{q_i^J\}$ over the volumes $\{\mathcal{V}_J\}$ of the corresponding objects $\{J\}$, as
\begin{equation}
F = -\beta^{-1} \ln \int_{\vecx_i^J \in \mathcal{V}_J} \prod_{i,J} d\vecx^J_i \, e^{-\beta H\left(\{\vecx_i^J\}\right)} \quad ,
\label{Z}
\end{equation}
where $\beta=1/(k_BT)$.
(The kinetic energy, $\sum_i p_i^2/2m_i$,  is easily integrated out of the partition function, and since it is independent of the arrangement of the objects, does not contribute to the force.) 
Charges $q_i^I$ and $q_j^J$ on different objects $I$ and $J$ interact via the Coulomb potential $q^I_i q^J_j G_M(\vecx^I_i,\vecx^J_j)$, where $G_M(\vecx,\vecx') =\left({4 \pi \ep_M  |\vecx-\vecx'|}\right)^{-1}$ is
the electrostatic Green's function for a medium with permittivity $\epsilon_M$, satisfying $-\ep_M\nabla^2 G_M(\vecx,\vecx') = \delta(\vecx-\vecx')$. 
The Hamiltonian, neglecting the kinetic energies, is thus,
\begin{equation}
H =
\sum_{I<J} \sum_{i,j}
q^I_i q^J_j G_M\left(\vecx^I_i,\vecx^J_j\right) +
\sum_{J} U_J\left(\{\vecx_i^J\}\right)\,,
\labeleqn{Hamiltonian}
\end{equation}
where the summation runs over $i\in I$, $j \in J$, and $U_J\left(\vecx_1^J,\vecx_2^J,\ldots\right)$ represents the interactions among the charges and with the `container' $J$. 
By choosing a high energetic cost of displacement some charges can be held stationary with respect to the container.

We can determine the stability of the objects' positions without explicit calculation of the partition function in Eq.~(\ref{Z}) by examining the change in free energy under an infinitesimal shift in the position of one object, while the others are held fixed. 
Under a translation of object $A$ by $\vecd$, the charges $q_1^A,\ldots,q_N^A$ in $A$ shift to positions $\vecx_1^A+\vecd,\ldots,\vecx_N^A+\vecd$. 
The interaction potential $U_A\left(\{\vecx_i^A\}\right)$, however, remains unchanged since the entire container $A$ is moved, that is, all vector quantities in $U_A\left(\{\vecx_i^A\}\right)$, including those specifying one-body potentials, are shifted by $+\vecd$. 
On the other hand, the Coulomb interaction between a charge $q_a^A$ in object $A$ and another charge $q_j^J$ in another object $J$ is modified to $q^A_a q^J_j G_M\left(\vecx^A_a+\vecd,\vecx^J_j\right)$. 
It is essential that the different objects {\em do not touch} to permit the infinitesimal translation of object $A$. The force on $A$ is given by $-\bnabla_\vecd F$.
The position of object $A$ is unstable if $\nabla_\vecd^2 F \leq 0$ and possibly stable if $\nabla^2_\vecd F > 0$.
%
%

The Laplacian of the free energy is given by
\be
\nabla^2_\vecd F
= \langle\nabla^2_\vecd H \rangle - \beta\left[ \langle\left(\bnabla_\vecd H \right)^2\rangle - \langle\bnabla_\vecd H \rangle^2 \right]\,, \ee where angular brackets denote thermal averages. The term in square brackets equals $\langle \left(\bnabla_\vecd H -\langle \bnabla_\vecd H\rangle \right)^2\rangle$, which is nonnegative and makes a destabilizing contribution.  
The Laplacian in the first average only acts on the Green's functions in \refeqn{Hamiltonian} which describe the interactions of the charges in object $A$ with charges in the other objects, e.g., $\left.\nabla_\vecd^2 G_M(\vecx^A_a+\vecd,\vecx^J_j) \right|_{\vecd=0}$; the $\delta$-functions resulting from such operations, e.g., $\delta(\vecx^A_a - \vecx^J_j)$, are always zero since the two charges lie in different volumes.  
Of course, the Laplacian of all other terms is zero since they do not depend on $\vecd$.  Thus, the result $\nabla_{\vecd}^2 F \leq 0$ follows principally from the vanishing of $\nabla^2_\vecd H$ for any configuration of charges within an object (as in the zero temperature Earnshaw case), and thermal fluctuations (second term) only enhance instability.

Upon requiring charge neutrality for objects in the previous model,  the fluctuating mobile charges generate fluctuating electric dipoles and higher multipoles on the objects. 
Even  at zero temperature, quantum charge and current fluctuations exists, generating Casimir forces.

Next, we proceed to  this quantum mechanical case by considering the stability of {\em neutral} objects interacting via the Casimir force, which can be attributed to either fluctuations of the EM field in the medium between the objects, or alternatively to fluctuating multipoles residing on them.
Explicit calculations for simple geometries indicate that the direction of the force can be predicted based on the relative permittivities, and permeabilities, of the objects and the medium.  Separating materials into two groups, with {\em (i)} permittivity higher than the medium or permeability lower than the medium ($\ep>\ep_M$ and $\mu\leq\mu_M$), or {\em (ii)} the other way around ($\ep<\ep_M$ and $\mu\geq\mu_M$), Casimir forces are found to be attractive between members of the same group and repulsive for different types.  
(While this has been shown in several examples, e.g. in Refs.~\cite{Dzyaloshinskii61,Feinberg68,Feinberg70,Boyer74,Kenneth02}, a theorem regarding the sign of the force only exists for mirror symmetric arrangements of objects~\cite{Kenneth06,Bachas07}.)  Since ordinary materials have permittivity higher than air and permeability very close to one, this effect causes objects to stick to one another. (The above statements will be made precise shortly.) Particularly for nanomachines this is detrimental as the Casimir force increases rapidly with decreasing separation.  This has motivated research into reversing the force; for example, a recent experiment~\cite{Munday09} shows that, in accord with the above rules, a dielectric medium can lead to repulsion. 
But the sign of the force is largely a matter of perspective, since attractive forces can be easily arranged to produce repulsion along a specific direction, e.g., as in Ref.~\cite{Rodriguez08}.  Instead, we focus on the question of stability which is more relevant to the design and development of MEMs and levitating devices.  We find that interactions between objects within the same class of material cannot produce stable configurations.  Thus, the force equilibria, as for example in Ref.~\cite{Schaden09}, are unstable since they are produced by geometric arrangements of infinite permittivity objects in vacuum.

Recent theoretical advances have led to new techniques, based on scattering theory, for efficiently computing the Casimir force (see, Ref.~\cite{Rahi09} for a detailed derivation, and a partial review of precursors~\cite{Emig07,Kenneth08,Maia_Neto08}).
The exact Casimir energy of an arbitrary number of objects with linear EM response, as described in the caption of \reffig{all} (right), is given by (see Eq.~(V.16) in Ref. \cite{Rahi09}),
\be
\mathcal{E} = \frac{\hbar c}{2\pi} \int_0^\infty d\ka\,\,\tr\ln \T^{-1} \T_\infty 
\, ,
\labeleqn{ECasimir}
\ee
where the operator $[\T^{-1}(ic\kappa,\vecx,\vecx')]$ equals
\be
\left(
\begin{array}{ c c c }
\left[\T_A^{-1}(ic\kappa,\vecx_1,\vecx'_1)\right] & \left[\tG     (ic\kappa,\vecx_1,\vecx'_2)\right] & \cdots \\
\left[\tG     (ic\kappa,\vecx_2,\vecx'_1)\right] & \left[\T_B^{-1}(ic\kappa,\vecx_2,\vecx'_2)\right] &  \\
\cdots   & & \cdots \\
\end{array}
\right)\,,
\ee
and $\T_\infty$ is the inverse of $\T^{-1}$ with $\tG$ set to zero. The square brackets ``[ ]'' denote the entire matrix or submatrix with rows indicated by $\vecx$ and columns by $\vecx'$.
To obtain the free energy $F$ at finite temperature, in place of the ground state energy $\mathcal{E}$,
 $\int \frac{d\ka}{2\pi}$ is replaced by the sum $\frac{kT}{\hbar c}\sum'_{\ka_n\geq 0}$ over Matsubara `wavenumbers,' $\ka_n = 2\pi n k T/\hbar c$, the primed summation implying that the $\kappa_0=0$ mode is weighted by $1/2$. 
 The operator $[\T^{-1}(ic\kappa,\vecx,\vecx')]$ has indices in position space. Each spatial index is limited to lie inside the objects $A,B,\cdots$. 
 For both indices $\vecx$ and $\vecx'$ in the same object $A$ the operator is just the inverse $\T$ operator of that object, $[\T_A^{-1}(ic\kappa,\vecx,\vecx')]$. 
For indices on different objects, $\vecx$ in $A$ and $\vecx'$ in $B$, it equals the electromagnetic Green's function operator $[\tG(ic\kappa,\vecx,\vecx')]$ for an isotropic, homogeneous medium.~\footnote{$\tG$ satisfies $\left(\curl\mu_M^{-1}(ic\kappa)\curl+\ep_M(ic\kappa)\kappa^2\right)\tG(ic\kappa,\vecx,\vecx')=\delta(\vecx-\vecx')\I$, and is related to $G_M$, the Green's function of the imaginary frequency Helmholtz equation, by $\tG(ic\ka,\vecx,\vecx') = \mu_M(ic\kappa)\left(\I + (n_M \ka)^{-2} \boldsymbol{\nabla}\otimes\boldsymbol{\nabla}'\right) G_M(icn_M\ka,\vecx,\vecx')$.~Here, $n_M(ic\ka)=\sqrt{\ep_M(ic\ka) \mu_M(ic\ka)}$ is the index of refraction of the medium, whose argument is suppressed to simplify the presentation. 
$G_M(ic\ka,\vecx,\vecx')=e^{-n_M\ka |\vecx-\vecx'|}/(4\pi|\vecx-\vecx'|)$ is the dynamic analogue of the electrostatic Green's function $G_M(\vecx,\vecx')$ in Eq.~\refeq{Hamiltonian}.}
As shown in Ref.~\cite{Rahi09}, after a few manipulations, the operators $\T_J$ and $\tG$ turn into the on-shell scattering amplitude matrix, $\mathbb{F}_J$, of object $J$ and the translation matrix $\X$, which converts wave functions between the origins of different objects. 
While practical computations require evaluation of the matrices in a particular basis, the basis independent operators $\T_J$ and $\tG$ are better suited to our general discussion here.

To investigate the stability of object $A$, we group the `rest' of the objects into a single entity $R$. 
So, $\T$ consists of $2 \times 2$ blocks, and the integrand in \refeqn{ECasimir} reduces to $\tr\ln\left(\I-\T_A\tG\T_R\tG\right)$. 
Merging the components of $R$ poses no conceptual difficulty given that the operators are expressed in a position basis, while an actual computation of the force between $A$ and $R$ would remain a daunting task.
If object $A$ is moved infinitesimally by vector $\vecd$, the Laplacian of the energy is given by
\begin{align}
\left.\nabla^2_{\vecd} \,\mathcal{E}\right|_{\vecd=0} & =
-\frac{\hbar c}{2\pi} \int_0^\infty d\ka \,\,\tr
\Big[
2 n_M^2(ic\kappa) \ka^2 \tfrac{\T_A \tG \T_R \tG}{\I-\T_A \tG \T_R \tG} \labeleqn{line1} \\
& + 2 \T_A \bnabla \tG \T_R \left( \bnabla \tG \right)^T \tfrac{\I}{\I-\T_A \tG \T_R \tG}  \labeleqn{line2} \\
& + 2 \T_A \bnabla \tG \T_R \tG \tfrac{\I}{\I-\T_A \tG \T_R \tG} \labeleqn{line3} \\
& \cdot
\left(\T_A \bnabla \tG \T_R \tG + \T_A \tG \T_R \left( \bnabla \tG \right)^T \right) \tfrac{\I}{\I-\T_A \tG \T_R \tG}  \Big] \nonumber \, .
\end{align}
After displacement of object $A$, the Green's function  multiplied by $\T_A$ on the left and $\T_R$ on the right $(\T_A\tG\T_R)$ becomes $\tG(ic\kappa,\vecx+\vecd,\vecx')$, while that multiplied by $\T_R$ on the left and $\T_A$ on the right $(\T_R\tG\T_A)$ becomes $\tG(ic\kappa,\vecx,\vecx'+\vecd)$. 
The two are related by transposition, and indicated by $\bnabla \tG(ic\kappa,\vecx,\vecx') = \bnabla_{\vecd} \tG(ic\kappa,\vecx+\vecd,\vecx')|_{\vecd=0}$ and $\left(\bnabla \tG(ic\kappa,\vecx,\vecx')\right)^T = \bnabla_{\vecd} \tG(ic\kappa,\vecx,\vecx'+\vecd)|_{\vecd=0}$ in the above equation.
In the first line we have substituted $n_M^2(ic\kappa) \kappa^2 \tG$ for $\nabla^2 \tG$ that differ only by derivatives of $\delta$--functions, which vanish since $\tG\left(ic\kappa,\vecx,\vecx'\right)$ is evaluated with $\vecx$ in one object and $\vecx'$ in another. 
In expressions not containing inverses of $\T$-operators, we can extend the domain of all operators to the entire space: 
$\T_J(ic\kappa,\vecx,\vecx')=0$ if $\vecx$ or $\vecx'$ are not on object $J$ and thus operator multiplication is unchanged. 

To determine the signs of the various terms to $\left.\nabla^2_{\vecd} \,\mathcal{E}\right|_{\vecd=0}$, we perform an analysis similar to Ref.~\cite{Kenneth06}. 
However, we do not investigate convergence issues and treat the operators like matrices from the start. This means that the necessary criteria (smoothness, boundedness, compact support, etc.) are assumed to be fulfilled in realistic situations, as dealt with in Ref.~\cite{Kenneth06}.
The operators $\T_J$ and $\tG$ are real and symmetric. 
An operator is positive (negative) semidefinite if all its eigenvalues are greater than or equal to zero (smaller than or equal to zero). 
It is easy to verify that $\tG$ is a positive semidefinite operator, since it is diagonal in momentum space, with $\tG(ic\ka,\veck) = \mu_M(ic\ka) \left(\I + \frac{\veck\otimes \veck}{n_M^2(ic\kappa) \kappa^2}\right)/\left(k^2+n_M^2(ic\kappa)\ka^2\right)$.
If $\mathbb{M}$ is a real and symmetric matrix, it is positive semidefinite if and only if there exists a matrix $\mathbb{B}$ such that $\mathbb{M}=\mathbb{B}^T \mathbb{B}$.
Let us assume that $\T_A$ and $\T_R$ are each either positive or negative semidefinite,
indicated by $s^A=\pm 1$ and $s^R=\pm 1$.
(We shall shortly show how the sign of $\T_J$ can be obtained from the object's permittivity and permeability.)
The eigenvalues of $\I-\T_A \tG \T_R \tG$, which equal those of $\I - s^A \sqrt{s^A \T_A} \tG \T_R \tG \sqrt{s^A \T_A}$, are strictly positive, since the energy is real. 
(The above expression appears in the integrand of \refeqn{ECasimir} if there are only two objects.)
Under the trace we always encounter the combination $\left(\I-\T_A\tG\T_R\tG\right)^{-1}\T_A$, which, taking advantage of its symmetries and definite sign, can be written as $s^A \mathbb{B}^T\mathbb{B}$, where $\mathbb{B}= \left(\I- s^A \sqrt{s^A \T_A} \tG \T_R \tG \sqrt{s^A \T_A}\right)^{-1/2} \sqrt{s^A \T_A}$.
The first term, line~\refeq{line1}, can now be rearranged as $\tr \, s^A \mathbb{B}^T\mathbb{B} \tG \T_R \tG = s^A s^R \tr \left[\left(\mathbb{B} \tG \mathbb{R}\right)\left(\mathbb{R}^T\tG\mathbb{B}^T\right)\right]$ by setting $\T_R=s^R \mathbb{R}\mathbb{R}^T$ and its sign is $s^A s^R$.
In the same way line~\refeq{line2} can be recast as $s^As^R \tr\left[\left(\mathbb{B}\bnabla\tG\mathbb{R}\right)\cdot\left(\mathbb{B}\bnabla\tG\mathbb{R}\right)^T\right]$, and its sign is thus also set by $s^A s^R$.
Lastly, the term in line \refeq{line3} can be rewritten as $\left(\mathbb{B} \bnabla \tG \T_R \tG \mathbb{B}^T + \mathbb{B} \tG \T_R \left(\bnabla \tG\right)^T \mathbb{B}^T\right)^2$. Since this is the square of a symmetric matrix, its eigenvalues are greater than or equal to zero, irrespective of the signs of $\T_A$ and $\T_R$. Overall, the Laplacian of the energy is smaller than or equal to zero as long as $s^A s^R \geq 0$.

What determines the sign of $\T_J$? 
It can be related to the electrodynamic `potential' $\V_J$, discussed in the next paragraph, by $\T_J \equiv \V_J/\left(\I+\tG \V_J\right)$~\cite{Rahi09}.
It is then positive or negative semidefinite depending on the sign $s^J$ of $\V_J$, since  $\T_J = s^J \sqrt{s^J \V_J}\frac{\I}{\I+ s^J \sqrt{s^J \V_J} \tG \sqrt{s^J \V_J}} \sqrt{s^J \V_J}$. 
The denominator $\I+ s^J \sqrt{s^J \V_J}\tG \sqrt{s^J \V_J}$ is positive semidefinite, even if $s^J=-1$, as its eigenvalues are the same as $\sqrt{\tG}(\tG^{-1}+\V_J)\sqrt{\tG}$; the term in the parantheses is just the (nonnegative) Hamiltonian of the field and the object $J$, $\tG^{-1}+\V_J = \curl\mu^{-1}(ic\ka,\vecx)\curl + \I \, \ka^2 \ep(ic\ka,\vecx)$. 
Here, we have used  $\V_J$ as given below, and $\ep(ic\ka,\vecx)$ and $\mu(ic\ka,\vecx)$ are the reponse functions defined everywhere in space, either of object $J$ or of the medium, depending on the point $\vecx$.

The analysis so far applies to each imaginary frequency $ic\ka$.  
As long as the signs of $\T_A$ and $\T_R$ are the same over the 
dominant frequencies in the integral (or the sum) in
\refeqn{ECasimir}, $\left.\nabla^2_{\vecd}
  \,\mathcal{E}\right|_{\vecd=0}\propto-s^As^R - (\text{positive
  term})$.  We are left to find the sign of the potential
$\mathbb{V}_J(ic\kappa,\vecx)=\I\,\ka^2\left(\ep_J(ic\ka,\vecx)-\ep_M(ic\ka)\right)
+ \curl\left(\mu_J^{-1}(ic\ka,\vecx)-\mu^{-1}_M(i\ka)\right)\curl$ of
the object $A$, and the compound object $R$
\footnote{The first curl in the operator $\mathbb{V}_J$ results from an integration by parts. It is understood that it acts on the wave function multiplying $\mathbb{V}_J$ from the left.}.
The sign is determined
by the relative permittivities and permeabilities of the objects and
the medium: If $\ep_J(ic\ka,\vecx) > \ep_M(ic\ka)$ and
$\mu_J(ic\ka,\vecx) \leq \mu_M(ic\ka)$ hold for all $\vecx$ in object
$J$, the potential $\V_J$ is positive.  If the opposite inequalities
are true, $\V_J$ is negative.  The curl operators surrounding the
magnetic permeability do not influence the sign, as in computing an
inner product with $\V_J$ they act symmetrically on both sides.  For
vacuum $\ep_M=\mu_M=1$, and material response functions
$\ep(ic\ka,\vecx)$ and $\mu(ic\ka,\vecx)$ are analytical continuations
of the permittivity and permeability for real
frequencies~\cite{LandauLifshitzEM84}.  While
$\ep(ic\ka,\vecx)> 1$ for positive $\ka$, there are no
restrictions other than positivity on $\mu(ic\ka,\vecx)$.
(For non-local and non-isotropic response, various inequalities must be generalized to the tensorial operators $\overleftrightarrow{\boldsymbol{\epsilon}}(ic\kappa,\vecx,\vecx')$ and $\overleftrightarrow{\boldsymbol{\mu}}(ic\kappa,\vecx,\vecx')$.)

In summary, if all objects fall into one of the two classes
described earlier, i) $\ep_J/\ep_M>1$ and $\mu_J/\mu_M\leq 1$ (positive
$\V_J$ and $\T_J$), or ii) $\ep_J/\ep_M<1$ and $\mu_J/\mu_M\geq 1$ with
(negative $\V_J$ and $\T_J$), none of the objects levitates stably.
(Under these conditions parallel slabs attract.)  In vacuum,
$\ep_M(ic\kappa)=\mu_M(ic\kappa)=1$; since $\ep(ic\ka,\vecx)> 1$ and
the magnetic response of ordinary materials is typically
negligible~\cite{LandauLifshitzEM84}, one concludes that
stable equilibria of the Casimir force do not exist.  If objects $A$
and $R$, however, belong to different categories --under which
conditions the parallel plate force is repulsive--, then the terms
under the trace in lines \refeq{line1} and \refeq{line2} are negative.
The positive term in line \refeq{line3} is typically smaller than the
first two, as it involves higher powers of $\T$ and $\tG$.  In this
case stable equilibrium is possible, as demonstrated recently for a
small inclusion within a dielectric filled cavity \cite{Rahi09-2}.
For the remaining two combinations of inequalities involving $\ep_J/\ep_M$ and $\mu_J/\mu_M$ the sign of $\V_J$ cannot be determined a priori. 
But for realistic distances between objects and the corresponding frequency ranges, the magnetic susceptibility is negligible for ordinary materials, and the inequalities involving $\mu$ can be ignored.

For levitation in vacuum one would need a strong paramagnet with negligible dielectric response. 
Metamaterials, incorporating arrays of micro-engineered circuity can display strong magnetic susceptibility at certain frequencies, and have been discussed as candidates for Casimir repulsion across vacuum~\cite{Rosa08,Rosa09}.
Given that such materials are fabricated out of ordinary metals and dielectrics with well-behaved $\ep(ic\kappa,\vecx)$ and $\mu(ic\kappa,\vecx)\approx 1$ at short scales, they should be subject
to the constraints above, ruling out the possibility of stable levitation.
Repulsion, in particular, is prohibited for such materials if one of the objects is an infinite slab with translational symmetry; as the energy as a function of separation from the slab would then have $\partial^2_d \mathcal{E}>0$ at some point since
the force has to vanish at infinite separation.
The above conclusions also apply to a massless scalar field, for which any arrangement of Dirichlet boundaries (corresponding to infinite permittivity here) is unstable to perturbations.

\begin{acknowledgments}
This research was supported by the NSF Grant No.  DMR-08-03315, and DARPA contract No. S-000354.
\end{acknowledgments}


\begin{thebibliography}{25}
\expandafter\ifx\csname natexlab\endcsname\relax\def\natexlab#1{#1}\fi
\expandafter\ifx\csname bibnamefont\endcsname\relax
  \def\bibnamefont#1{#1}\fi
\expandafter\ifx\csname bibfnamefont\endcsname\relax
  \def\bibfnamefont#1{#1}\fi
\expandafter\ifx\csname citenamefont\endcsname\relax
  \def\citenamefont#1{#1}\fi
\expandafter\ifx\csname url\endcsname\relax
  \def\url#1{\texttt{#1}}\fi
\expandafter\ifx\csname urlprefix\endcsname\relax\def\urlprefix{URL }\fi
\providecommand{\bibinfo}[2]{#2}
\providecommand{\eprint}[2][]{\url{#2}}

\bibitem[{\citenamefont{Earnshaw}(1842)}]{Earnshaw42}
\bibinfo{author}{\bibfnamefont{S.}~\bibnamefont{Earnshaw}},
  \bibinfo{journal}{Trans. Camb. Phil. Soc.} \textbf{\bibinfo{volume}{7}},
  \bibinfo{pages}{97} (\bibinfo{year}{1842}).

\bibitem[{\citenamefont{Geim}(1998)}]{Geim98}
\bibinfo{author}{\bibfnamefont{A.}~\bibnamefont{Geim}}, \bibinfo{journal}{Phys.
  Today} \textbf{\bibinfo{volume}{51(9)}}, \bibinfo{pages}{36}
  (\bibinfo{year}{1998}).

\bibitem[{\citenamefont{Braunbek}(1939{\natexlab{a}})}]{Braunbek39-1}
\bibinfo{author}{\bibfnamefont{W.}~\bibnamefont{Braunbek}},
  \bibinfo{journal}{Z. Phys.} \textbf{\bibinfo{volume}{112}},
  \bibinfo{pages}{753} (\bibinfo{year}{1939}{\natexlab{a}}).

\bibitem[{\citenamefont{Braunbek}(1939{\natexlab{b}})}]{Braunbek39-2}
\bibinfo{author}{\bibfnamefont{W.}~\bibnamefont{Braunbek}},
  \bibinfo{journal}{Z. Phys.} \textbf{\bibinfo{volume}{112}},
  \bibinfo{pages}{764} (\bibinfo{year}{1939}{\natexlab{b}}).

\bibitem[{\citenamefont{Jones}(1995)}]{Jones95}
\bibinfo{author}{\bibfnamefont{T.~B.} \bibnamefont{Jones}},
  \emph{\bibinfo{title}{Electromechanics of Particles}}
  (\bibinfo{publisher}{Cambridge University Press},
  \bibinfo{address}{Cambridge}, \bibinfo{year}{1995}).

\bibitem[{\citenamefont{Ashkin}(1970)}]{Ashkin70}
\bibinfo{author}{\bibfnamefont{A.}~\bibnamefont{Ashkin}},
  \bibinfo{journal}{Phys. Rev. Lett.} \textbf{\bibinfo{volume}{24}},
  \bibinfo{pages}{156} (\bibinfo{year}{1970}).

\bibitem[{\citenamefont{Ashkin and Gordon}(1983)}]{Ashkin83}
\bibinfo{author}{\bibfnamefont{A.}~\bibnamefont{Ashkin}} \bibnamefont{and}
  \bibinfo{author}{\bibfnamefont{J.~P.} \bibnamefont{Gordon}},
  \bibinfo{journal}{Opt. Lett.} \textbf{\bibinfo{volume}{8}},
  \bibinfo{pages}{511} (\bibinfo{year}{1983}).

\bibitem[{\citenamefont{Dzyaloshinskii
  et~al.}(1961)\citenamefont{Dzyaloshinskii, Lifshitz, and
  Pitaevskii}}]{Dzyaloshinskii61}
\bibinfo{author}{\bibfnamefont{I.~E.} \bibnamefont{Dzyaloshinskii}},
  \bibinfo{author}{\bibfnamefont{E.~M.} \bibnamefont{Lifshitz}},
  \bibnamefont{and} \bibinfo{author}{\bibfnamefont{L.~P.}
  \bibnamefont{Pitaevskii}}, \bibinfo{journal}{Adv. Phys.}
  \textbf{\bibinfo{volume}{10}}, \bibinfo{pages}{165} (\bibinfo{year}{1961}).

\bibitem{Feinberg68}
G.~Feinberg and J.~Sucher,~J.~Chem.~Phys.,~\textbf{48},~3333~(1968).

\bibitem[{\citenamefont{Feinberg and Sucher}(1970)}]{Feinberg70}
\bibinfo{author}{\bibfnamefont{G.}~\bibnamefont{Feinberg}} \bibnamefont{and}
  \bibinfo{author}{\bibfnamefont{J.}~\bibnamefont{Sucher}},
  \bibinfo{journal}{Phys. Rev. A} \textbf{\bibinfo{volume}{2}},
  \bibinfo{pages}{2395} (\bibinfo{year}{1970}).

\bibitem[{\citenamefont{Boyer}(1974)}]{Boyer74}
\bibinfo{author}{\bibfnamefont{T.~H.} \bibnamefont{Boyer}},
  \bibinfo{journal}{Phys. Rev. A} \textbf{\bibinfo{volume}{9}},
  \bibinfo{pages}{2078} (\bibinfo{year}{1974}).

\bibitem[{\citenamefont{Kenneth et~al.}(2002)\citenamefont{Kenneth, Klich,
  Mann, and Revzen}}]{Kenneth02}
\bibinfo{author}{\bibfnamefont{O.}~\bibnamefont{Kenneth}},
  \bibinfo{author}{\bibfnamefont{I.}~\bibnamefont{Klich}},
  \bibinfo{author}{\bibfnamefont{A.}~\bibnamefont{Mann}}, \bibnamefont{and}
  \bibinfo{author}{\bibfnamefont{M.}~\bibnamefont{Revzen}},
  \bibinfo{journal}{Phys. Rev. Lett.} \textbf{\bibinfo{volume}{89}},
  \bibinfo{pages}{033001} (\bibinfo{year}{2002}).

\bibitem[{\citenamefont{Kenneth and Klich}(2006)}]{Kenneth06}
\bibinfo{author}{\bibfnamefont{O.}~\bibnamefont{Kenneth}} \bibnamefont{and}
  \bibinfo{author}{\bibfnamefont{I.}~\bibnamefont{Klich}},
  \bibinfo{journal}{Phys. Rev. Lett.} \textbf{\bibinfo{volume}{97}},
  \bibinfo{eid}{160401} (\bibinfo{year}{2006}).

\bibitem[{\citenamefont{Bachas}(2007)}]{Bachas07}
\bibinfo{author}{\bibfnamefont{C.~P.} \bibnamefont{Bachas}},
  \bibinfo{journal}{J. Phys. A: Math. Theor.} \textbf{\bibinfo{volume}{40}},
  \bibinfo{pages}{9089} (\bibinfo{year}{2007}).

\bibitem[{\citenamefont{Munday et~al.}(2009)\citenamefont{Munday, Capasso, and
  Parsegian}}]{Munday09}
\bibinfo{author}{\bibfnamefont{J.~N.} \bibnamefont{Munday}},
  \bibinfo{author}{\bibfnamefont{F.}~\bibnamefont{Capasso}}, \bibnamefont{and}
  \bibinfo{author}{\bibfnamefont{V.~A.} \bibnamefont{Parsegian}},
  \bibinfo{journal}{Nature} \textbf{\bibinfo{volume}{457}},
  \bibinfo{pages}{170} (\bibinfo{year}{2009}).

\bibitem[{\citenamefont{Rodriguez et~al.}(2008)\citenamefont{Rodriguez,
  Joannopoulos, and Johnson}}]{Rodriguez08}
\bibinfo{author}{\bibfnamefont{A.~W.} \bibnamefont{Rodriguez}},
  \bibinfo{author}{\bibfnamefont{J.~D.} \bibnamefont{Joannopoulos}},
  \bibnamefont{and} \bibinfo{author}{\bibfnamefont{S.~G.}
  \bibnamefont{Johnson}}, \bibinfo{journal}{Phys. Rev. A}
  \textbf{\bibinfo{volume}{77}}, \bibinfo{eid}{062107} (\bibinfo{year}{2008}).

\bibitem[{\citenamefont{Schaden}(2009)}]{Schaden09}
\bibinfo{author}{\bibfnamefont{M.}~\bibnamefont{Schaden}},
  \bibinfo{journal}{Phys. Rev. Lett.} \textbf{\bibinfo{volume}{102}},
  \bibinfo{eid}{060402} (\bibinfo{year}{2009}).

\bibitem[{\citenamefont{Rahi et~al.}(2009)\citenamefont{Rahi, Emig, Graham,
  Jaffe, and Kardar}}]{Rahi09}
\bibinfo{author}{\bibfnamefont{S.~J.} \bibnamefont{Rahi}},
  \bibinfo{author}{\bibfnamefont{T.}~\bibnamefont{Emig}},
  \bibinfo{author}{\bibfnamefont{N.}~\bibnamefont{Graham}},
  \bibinfo{author}{\bibfnamefont{R.~L.} \bibnamefont{Jaffe}}, \bibnamefont{and}
  \bibinfo{author}{\bibfnamefont{M.}~\bibnamefont{Kardar}},
  \bibinfo{journal}{Phys. Rev. D} \textbf{\bibinfo{volume}{80}},
  \bibinfo{eid}{085021} (\bibinfo{year}{2009}).

\bibitem[{\citenamefont{Emig et~al.}(2007)\citenamefont{Emig, Graham, Jaffe,
  and Kardar}}]{Emig07}
\bibinfo{author}{\bibfnamefont{T.}~\bibnamefont{Emig}},
  \bibinfo{author}{\bibfnamefont{N.}~\bibnamefont{Graham}},
  \bibinfo{author}{\bibfnamefont{R.~L.} \bibnamefont{Jaffe}}, \bibnamefont{and}
  \bibinfo{author}{\bibfnamefont{M.}~\bibnamefont{Kardar}},
  \bibinfo{journal}{Phys. Rev. Lett.} \textbf{\bibinfo{volume}{99}},
  \bibinfo{pages}{170403} (\bibinfo{year}{2007}).

\bibitem[{\citenamefont{Kenneth and Klich}(2008)}]{Kenneth08}
\bibinfo{author}{\bibfnamefont{O.}~\bibnamefont{Kenneth}} \bibnamefont{and}
  \bibinfo{author}{\bibfnamefont{I.}~\bibnamefont{Klich}},
  \bibinfo{journal}{Phys. Rev. B} \textbf{\bibinfo{volume}{78}},
  \bibinfo{eid}{014103} (\bibinfo{year}{2008}).

\bibitem[{\citenamefont{{Maia Neto} et~al.}(2008)\citenamefont{{Maia Neto},
  Lambrecht, and Reynaud}}]{Maia_Neto08}
\bibinfo{author}{\bibfnamefont{P.~A.} \bibnamefont{{Maia Neto}}},
  \bibinfo{author}{\bibfnamefont{A.}~\bibnamefont{Lambrecht}},
  \bibnamefont{and} \bibinfo{author}{\bibfnamefont{S.}~\bibnamefont{Reynaud}},
  \bibinfo{journal}{Phys. Rev. A} \textbf{\bibinfo{volume}{78}},
  \bibinfo{eid}{012115} (\bibinfo{year}{2008}).

\bibitem[{\citenamefont{Landau and Lifshitz}(1984)}]{LandauLifshitzEM84}
\bibinfo{author}{\bibfnamefont{L.~D.} \bibnamefont{Landau}} \bibnamefont{and}
  \bibinfo{author}{\bibfnamefont{E.~M.} \bibnamefont{Lifshitz}},
  \emph{\bibinfo{title}{Electrodynamics of continuous media}}
  (\bibinfo{publisher}{Pergamon Press}, \bibinfo{address}{Oxford},
  \bibinfo{year}{1984}).

\bibitem[{\citenamefont{Rahi and Zaheer}(2009)}]{Rahi09-2}
\bibinfo{author}{\bibfnamefont{S.~J.} \bibnamefont{Rahi}} \bibnamefont{and}
  \bibinfo{author}{\bibfnamefont{S.}~\bibnamefont{Zaheer}}
  (\bibinfo{year}{2009}), \eprint{arXiv:0909.4510}.

\bibitem[{\citenamefont{Rosa et~al.}(2008)\citenamefont{Rosa, Dalvit, and
  Milonni}}]{Rosa08}
\bibinfo{author}{\bibfnamefont{F.~S.~S.} \bibnamefont{Rosa}},
  \bibinfo{author}{\bibfnamefont{D.~A.~R.} \bibnamefont{Dalvit}},
  \bibnamefont{and} \bibinfo{author}{\bibfnamefont{P.~W.}
  \bibnamefont{Milonni}}, \bibinfo{journal}{Phys. Rev. Lett.}
  \textbf{\bibinfo{volume}{100}}, \bibinfo{eid}{183602} (\bibinfo{year}{2008}).

\bibitem[{\citenamefont{Rosa}(2009)}]{Rosa09}
\bibinfo{author}{\bibfnamefont{F.~S.~S.} \bibnamefont{Rosa}},
  \bibinfo{journal}{J. Phys.: Conf. Ser.} \textbf{\bibinfo{volume}{161}},
  \bibinfo{pages}{012039} (\bibinfo{year}{2009}).

\end{thebibliography}
\end{document}